\documentclass{ws-procs9x6}

\usepackage{graphicx}
\usepackage{amssymb}

\usepackage{bm}

\begin{document}

\title{Stochastic gauge: a new technique for quantum simulations}

\author{Peter Drummond, P. Deuar, J. F. Corney, K. Kheruntsyan\\
 \textit{Australian Centre for Quantum Atom Optics,}\\
 \textit{University of Queensland, AUSTRALIA}}

\maketitle
\abstracts{We review progress towards direct simulation of quantum dynamics in many-body systems, using recently developed
stochastic gauge techniques. We consider master equations, canonical ensemble calculations and reversible quantum dynamics
are compared, as well the general question of strategies for choosing the gauge.
}

\section{INTRODUCTION}

In this paper, we will review progress toward the direct simulation of quantum dynamics in many-body systems. This is a central
problem in modern theoretical physics, sometimes claimed to be inherently insoluble\cite{Feynman} on digital computers,
due to complexity issues. In particular, we will evaluate progress and results using stochastic gauge methods\cite{GaugeP}
on currently available digital computers.

This is a timely issue to the spectroscopist at the start of the twenty-first century, since spectroscopy now involves the
study of increasingly complex systems - not just molecules, but large interacting many-body systems like dilute bosonic and
fermionic quantum gases. The use of subtle dynamical experiments allows the exploration of many-body dynamics well beyond
regimes where perturbation theory is useful, and experiments in regimes where mean-field theory is not applicable are increasingly
common.

The specific issues treated are:

\begin{itemize}
\item Stochastic gauge \emph{}- this is a unified simulation method which provides a route to understanding a wide range of stochastic
methods.
\item Master equations - this is the most general real time problem, which can include both reversible and/or dissipative elements.
\item Canonical simulations - needed if a finite temperature grand canonical ensemble is to be calculated, in thermal equilibrium.
\item Quantum dynamics \emph{}- we illustrate this by a calculation of correlated atom pairs in coherent molecular conversion to
atoms. 
\end{itemize}

\subsection{The problem: complexity}

Before giving an account of these new techniques, we must consider what the problem is. As an example, consider modern BEC
experiments, which may involve \textbf{}$10^{100000}$ states, or around $10^{6}$ equivalent qubits in Hilbert space. Such
many-particle dynamical experiments are commonplace, but they do place severe demands on theory. 

Apart from various approximate methods, direct computation in a number-state basis is useful only for small numbers of particles;
in general, it needs exponentially too much memory! The last problem led to a famous and somewhat pessimistic conclusion
by Feynman\cite{Feynman}, summarized in the following statement: \emph{{}``Can a quantum system be probabilistically simulated
by a classical universal computer? ..the answer is certainly,} \textbf{\emph{No}}\emph{!''}

\subsection{Quantum Computers: too small, too costly?}

A possible solution to the complexity problem was suggested by Feynman. This is the proposal to develop quantum computers
in which the logical entities are qubits - that is, two-state quantum systems - and the logical operations are dissipation-free
unitary transformations on large numbers of qubits. This solves the quantum complexity problem at the software level - since
qubits are extremely efficient at storing large Hilbert spaces - but there is still a hardware issue to be solved. 

Great progress has been made at NIST\cite{Wineland} (also reported at this conference), in constructing ion trap devices
that are able to carry out binary qubit logic operations. Despite this, even if the NIST ion traps are able to be scaled
to the projected size of $10^{4}$ ions, they will still be relatively small due to error-correction overheads, and also
rather slow and expensive. This is illustrated in the following table.

\begin{table}
\begin{tabular}{|c|c|c|c|c|c|}
\hline 
COMPUTER&
 bits&
 Hz&
 Ops/cycle&
 \$/CPU&
Ops/s/\$\tabularnewline
\hline
DIGITAL(2000)&
 $10^{10}$&
 $10^{9}$&
 $10^{2}$&
 $10^{3}$&
 $10^{8}$\tabularnewline
\hline
QC (2000)&
 $10$&
 $10^{3}$&
 $1$&
 $10^{6}$&
 $10^{-3}$\tabularnewline
\hline
DIGITAL (2020)&
 $10^{12}$&
 $10^{10}$&
 $10^{3}$&
 $10$&
 $10^{12}$\tabularnewline
\hline
QC (2020)&
 $10^{2}$&
 $10^{4}$&
 $10$&
 $10^{5}$&
 $1$ \tabularnewline
\hline
\end{tabular}

\caption{Estimates of digital vs quantum computer performance}
\end{table}

While the digital computer projections are industry-based\cite{Digital}, the quantum computer projections are optimistic
in assuming that all of the known problems will be solved on this timescale.

\section{Phase-space representations}

There is a possible alternative: improved digital algorithms that may allow industry standard digital computers to be used
for quantum simulations. The proposal treated here is to map quantum dynamics into equations for stochastic motion on a phase-space,
which are then randomly sampled to obtain statistical averages of quantum observables. The procedure is similar to path-integral
Monte-Carlo techniques\cite{wilson:74} used for canonical ensembles.

Historically, this method originated in the classical phase-space methods: the \textbf{Wigner\cite{Wig-Wigner}, Q}\cite{Hus-Q},
and \textbf{P\cite{Gla-P}} representations all have a classical phase-space dimension of (say) $d$ dimensions. However,
the quantum dynamics of systems of \emph{interacting} particles cannot be sampled stochastically. More recently, methods
of larger dimension using `quantum' phase-spaces were introduced: the first was the \textbf{Positive-P} representation\cite{+P},
using a generalized phase-space variable $\bm\alpha$ with $d$ complex dimensions. This allows dynamical problems to be
mapped into stochastic equations, given some restrictions on the tails of the distribution. 

The positive-P technique has been used widely in quantum optics, and led to the prediction of quantum soliton squeezing\cite{carterdrummond:87},
which was verified experimentally\cite{Solexp}. While the first calculations used linearization, these were quickly extended\cite{carterdrummond:87}
to the first real-time quantum field simulations with $10^{9}$ particles in $10^{3}$ modes, corresponding to a Hilbert
space dimension of more than $10^{4}$ qubits. 

More recently, these methods were extended to many other nonlinear quantum systems, including full two and three dimensional
calculations, typically with damping present. An example of this is the first a-priori quantum simulation\cite{evapcool}
of the formation of a BEC via evaporative cooling. With $2\times10^{4}$ particles in $3\times10^{4}$ 
modes, this corresponds
to over $10^{5}$ qubits.

Despite its success, the basic positive-P technique does have limitations. The stochastic equations have stability problems
if there is not sufficient damping in the system, leading to uncontrollable growth in sampling error after a certain simulation
time. Systematic `boundary term' errors\cite{GGD-Validity} can arise for certain nonlinear systems.

Both of these problems can be solved by use of \textbf{stochastic gauges}\cite{GaugeP}. This extension of the positive-P
technique introduces an extra phase-space variable $\Omega$, which behaves as a weighting function, and allows the introduction
of arbitrary `drift gauges' $\mathbf{g}(\bm\alpha)$, which can be used to stabilize stochastic paths without affecting the
physical ensemble averages. In addition, there are `diffusion gauges' $\mathbf{g}^{d}(\bm\alpha)$, which exploit a freedom
in choice of noise terms to reduce the sampling error. The stochastic gauge method is a very general technique, which also
unifies some earlier methods\cite{Paris1,Plimak}. Its flexibility enables it to be tailored to different applications.

\subsection{Method in outline}

A general quantum calculation in real or imaginary time (for thermal equilibrium) can be written as a Liouville equation
for the density operator:\begin{eqnarray}
\partial\widehat{\rho}(t)/\partial t & = & \widehat{L}[\widehat{\rho}(t)].\end{eqnarray}
 To solve this with stochastic gauges, we first expand the density operator over some suitable operator basis $\widehat{\Lambda}(\overrightarrow{\lambda})$:\begin{eqnarray}
\widehat{\rho}(t) & = & \int P(\overrightarrow{\lambda},t)\widehat{\Lambda}(\overrightarrow{\lambda})d^{2(d+1)}\overrightarrow{\lambda},\end{eqnarray}
 which defines a $(d+1)$-dimensional complex phase-space: $\overrightarrow{\lambda}=(\Omega\,,\bm\alpha)$. For a given
basis, there are identities which allow us to write the Liouville operator equation as\begin{eqnarray}
\partial\widehat{\rho}(t)/\partial t & = & \int P(\overrightarrow{\lambda},t)\mathcal{L}_{A}\widehat{\Lambda}(\overrightarrow{\lambda})d^{2(d+1)}\overrightarrow{\lambda},\end{eqnarray}
where $\mathcal{L}_{A}$ is a linear differential operator which can include arbitrary gauge functions. If there are no derivatives
higher than second order, this equation can be transformed to a Fokker-Planck equation for $P$, provided the gauges are
chosen to eliminate any boundary terms that may otherwise arise from the partial integration. This can always be transformed
into the stochastic equations:\begin{eqnarray}
d\Omega/\partial t & = & \Omega\left[U+\,\mathbf{g}\,\cdot\bm\zeta\right]\label{eq:gauge1}\\
d\bm\alpha/\partial t & = & \mathbf{A}+\mathbf{B}(\bm\zeta-\,\mathbf{g}),\label{eq:gauge2}\end{eqnarray}
 where $U$ and the vector $\mathbf{A}$ are determined by the form of the original Liouville equation. The drift gauges
appear as the arbitrarily functions $\mathbf{g}$, and diffusion gauges appear as the freedom that exists in choosing the
noise matrix $\mathbf{B}$. The noise terms $\bm\zeta$ are Gaussian white noises, with correlations:\begin{equation}
\langle\zeta_{i}(t)\zeta_{j}(t')\rangle=\delta_{ij}\delta(t-t').\label{eq:gauge3}\end{equation}

Equations (\ref{eq:gauge1}-\ref{eq:gauge3}) form the central result of the stochastic gauge method and can be used to solve
a large class of quantum dynamical and thermal-equilibrium problems\cite{GaugeP}. 

In practice, the numerical implementation of these equations is simplified by use of an extensible multi-dimensional simulator (XMDS) which generates efficient C++ computer code 
from a brief high-level problem description. It also
features automatic clustering with standard message-passing interface (MPI) routines, and is free under the open-source GNU public license (GPL) at the website \emph{www.xmds.org}.

There are three main types of problems which can be studied using these methods: 1) master equations of open quantum systems,
which feature damping as well as coherent evolution, 2) canonical ensembles, which involve the `imaginary time' calculation
of thermal equilibrium states and 3) quantum dynamics of closed many-body systems. We now present an example of each type
that has been solved by use of stochastic gauge methods.

\section{Master Equation}

Accurate simulation of experimental systems usually requires taking into account damping and other coupling to the external
environment, which can be achieved with a master equation. The positive-P method has been shown to suffer from boundary term
systematic errors\cite{GGD-Validity} in some nonlinear cases when there is small linear damping, for example in a coherently
driven interferometer with two-boson damping. Introducing appropriate drift stochastic gauges removes these systematic errors.

A typical master equation , when reduced to a single mode, can be written \begin{eqnarray}
\frac{\partial\hat{\rho}}{\partial t} & = & \left[\varepsilon\hat{a}^{\dagger}-\varepsilon^{*}\hat{a},\hat{\rho}\right]+\frac{1}{2}(2\hat{a}^{2}\hat{\rho}\hat{a}^{\dagger2}-\hat{a}^{\dagger2}\hat{a}^{2}\hat{\rho}-\hat{\rho}\hat{a}^{\dagger2}\hat{a}^{2})\,\,,\end{eqnarray}
 with $\varepsilon$ the driving field amplitude. The performance of stochastic gauge simulations has been compared to positive-P
and exact results in earlier work\cite{GaugeP}. Due to space limitations, we refer the reader to this article, which shows
that in all the cases studied, any systematic errors are removed and the useful simulation time is greatly extended. The
stochastic gauge method then becomes the preferred approach when applied to a many-particle or many-mode system in which
a direct number-state basis calculation is impractical.

\section{Grand canonical ensembles}

We consider the equilibrium states of an interacting Bose gas, thermally and diffusively coupled to a reservoir. The calculation
of spatial correlations in a one (or more) -dimensional interacting Bose gas is a highly non-trivial task, and has been calculated
for the first time using the stochastic gauge method\cite{Paris1,pdpdcanprl}. The positive-P method is inadequate to the
task because of boundary term errors, and the inability to handle the Gibbs weight factors.

The imaginary time evolution of grand canonical ensembles (with temperature $T$) can be modeled using an anti-commutator equation of form: 
\begin{equation}
\frac{\partial\hat{\rho}_{u}}{\partial\tau}=\left[\widehat{N}\frac{\partial\mu}{\partial\tau}-\widehat{H},\hat{\rho}_{u}\right]_{+}.\end{equation}
 Here, $\tau=1/k_{B}T$, $\mu$ is the chemical potential, $\widehat{N}$ is particle number, $\widehat{H}$ the Hamiltonian,
and $\hat{\rho}_{u}$ is un-normalized. One starts at very high temperature $\tau=0$, where the ensemble is known, and evolves
in $\tau$ to get finite-temperature ensembles.

We model the 1D Bose gas with the Bose-Hubbard Hamiltonian ($\hbar=1$)\begin{equation}
\widehat{H}=\left[\sum_{i}\sum_{i}\omega_{ij}\hat{a}_{i}^{\dagger}\hat{a}_{j}+\sum_{j}:\widehat{n}_{j}^{2}:\right]\,\end{equation}
 on a lattice labeled by $i$. The $\omega_{ij}$ are nonlocal couplings between modes, and $\widehat{n}_{i}$ are occupations
at each point.

One obtains the equations \begin{eqnarray}
\frac{d\alpha_{i}}{d\tau} & = & -[n_{i}+ig_{i}-i\zeta_{i}(\tau)]\alpha_{i}-\sum_{j}\widetilde{\omega}_{ij}\alpha_{j}/2,\nonumber \\
\frac{d\beta_{i}}{d\tau} & = & -[n_{i}+ig_{i}-i\widetilde{\zeta}_{i}(\tau)]\beta_{i}-\sum_{j}\widetilde{\omega}_{ji}\beta_{j}/2,\nonumber \\
\frac{d\Omega}{d\tau} & = & -\sum_{ij}\widetilde{\omega}_{ij}\beta_{i}\alpha_{j}-\sum_{i}n_{i}^{2}+\sum_{i}g_{i}[\zeta_{i}(\tau)+\widetilde{\zeta}_{i}(\tau)].\label{gaugeequn}\end{eqnarray}
 Here the $\widetilde{\omega}_{ij}$ include $\omega_{ij}$ together with terms coming from the chemical potential, and $n_{i}=\alpha_{i}\beta_{i}$,
while $\zeta_{i}(\tau)$ and $\widetilde{\zeta}_{i}(\tau)$ are all independent Gaussian noises of variance $\delta(\tau)$.
We choose the stabilizing gauge $g_{i}=i(\textrm{Re}[n_{i}]-|n_{i}|)$ to remove boundary term errors, and reduce sampling
error.

\begin{figure}
\includegraphics[%
  width=10cm]{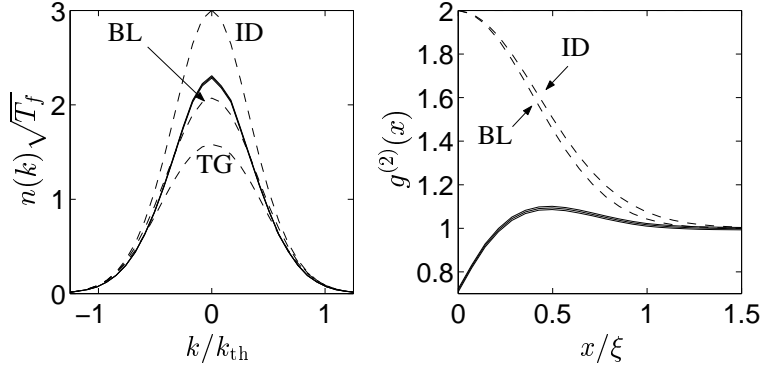}

\caption{\label{g2xFIG} Momentum distribution $n(k)$ and correlation function $g^{(2)}(x)$ of uniform 1-D Bose gas (solid lines).
Here $\gamma=10=mg/\rho$ is the scaled interaction strength (with $\rho$ the linear density), and $T=10T_{d}$, where $T_{d}=2\pi\rho^{2}/mk_{B}$
is the quantum degeneracy temperature and $\xi$ is the healing length. ID is for an ideal gas ($\gamma=0$ and $T=10T_{d}$),
while BL is for a Boltzmann gas ($\gamma=10$, $T/T_{d}\rightarrow\infty$). }
\end{figure}

The second-order correlation function \begin{equation}
g^{(2)}(x)=\langle:\widehat{n}(0)\widehat{n}(x):\rangle/\langle\widehat{n}(0)\rangle\langle\widehat{n}(x)\rangle\end{equation}
 is shown in Fig.~\ref{g2xFIG} for a uniform one-dimensional Bose gas in the Tonks-Girardeau regime of incipient fermionization. 

It is seen that $g^{(2)}(x)$ is well calculated, and shows that in this parameter regime, there is a preferred inter-atom
distance of approximately half the healing length. Anti-bunching ($g^{(2)}(0)=0.72\pm0.01$) is also seen. There is complete
agreement with all known finite temperature exact results\cite{1Dg2}.

\section{Quantum Dynamics}

Here we consider the process of coherent dissociation of a BEC of molecular dimers into pairs of constituent atoms, via stimulated
Raman transitions or Feshbach resonances\cite{Twinbeams}. The resulting effect is formation of quantum correlated twin atom-laser
beams with squeezing in the particle number difference. This is the matter wave analog of optical parametric down-conversion
producing the famous entangled photon pairs.

The effective Hamiltonian interaction, taken for simplicity in one space dimension, is \begin{equation}
\hat{H}^{(\chi)}=i\frac{\chi(t)}{2}\int dx\left[e^{-i\omega t}\hat{\Psi}_{2}\hat{\Psi}_{1}^{\dagger\,2}-e^{i\omega t}\hat{\Psi}_{2}^{\dagger}\hat{\Psi}_{1}^{2}\right]\,\,.\end{equation}
 Here, $\hat{\Psi}_{1}(x,t)$ and $\hat{\Psi}_{2}(x,t)$ are the atomic and molecular field operators, respectively, $\,\chi(t)$
is the atom-molecule coupling responsible for the conversion of molecules into atom pairs (and vice versa), and $\omega$
is a detuning. The complete Hamiltonian also contains the usual kinetic energy terms, external trapping potentials, as well
as usual quartic interaction terms describing atom-atom, atom-molecule, and molecule-molecule $s$-wave scattering.

The quantum dynamical simulation of this process is carried out using the $+P$ representation. This is an example where
interesting physical effects can take place on short time scales so that the straightforward $+P$ simulations give reliable
results.

For short dissociation time intervals such that the atomic self-interaction can be neglected due to low atomic density, and
for large effective detuning such that the atom-molecule $s$-wave interaction can be neglected too, the dynamics is simulated
via the following set of stochastic equations (in appropriate dimensionless units): \begin{eqnarray}
\frac{\partial\psi_{1}}{\partial\tau} & = & i\frac{\partial^{2}\psi_{1}}{\partial\xi^{2}}-(\gamma+i\delta)\psi_{1}+\kappa\psi_{2}\psi_{1}^{+}+\sqrt{\kappa\psi_{2}}\eta_{1},\nonumber \\
\frac{\partial\psi_{2}}{\partial\tau} & = & \frac{i}{2}\frac{\partial^{2}\psi_{2}}{\partial\xi^{2}}-iv(\xi,\tau)\psi_{2}-\frac{\kappa}{2}\psi_{1}^{2}+\sqrt{-iu}\psi_{2}\eta_{2},\end{eqnarray}
 together with the corresponding {}``conjugate'' equations for the $\psi_{1,2}^{+}$-fields. Here, $\xi$ and $\tau$ are
the dimensionless coordinate and time, $\kappa$ is the coupling proportional to $\chi$, $v(\xi,\tau)$ accounts for the
molecular trapping potential and $u$ the molecule-molecule self-interaction, $\delta$ is the dimensionless detuning parameter,
$\gamma$ accounts for possible atomic losses, and $\eta_{1,2\,}$ are the Gaussian delta-correlated noise terms.

\begin{figure}
\begin{center} \includegraphics[  width=10cm]
{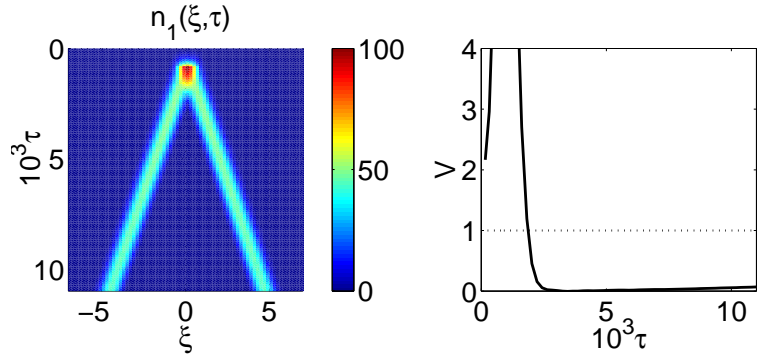}\end{center}

\caption{Evolution of the atomic density (left panel) and the variance $V$ versus time (right panel). The dissociation takes place
in the time interval from $\tau=0$ to $8\times10^{-4}$, followed by free evolution of the atomic beams.}

\label{twins}
\end{figure}

A typical example demonstrating the formation of two counter-propagating atomic beams is given in Fig. \ref{twins}. The
correlations between the beams propagating to the left (-) and to the right (+) are quantified via the normalized variance
of fluctuations in the relative total number of particles, $V=\left\langle [\Delta(\hat{N}_{-}-\hat{N}_{+})]^{2}\right\rangle \;/\;(\left\langle \hat{N}_{-}\right\rangle +\left\langle \hat{N}_{+}\right\rangle )$.
In a coherent (uncorrelated) state, the variance is $V=1$, while $V<1$ implies squeezing of quantum fluctuations below
the coherent level -- which is due to strong quantum correlations between the atoms in the two beams. The example represented
in Fig. \ref{twins} shows more than $93\%$ squeezing, once the dissociation is switched off and the atomic beams separate
spatially.

The physical reason for the correlation and the squeezing is momentum conservation, which requires that each atom emitted
with a (dimensionless) momentum $q>0$ be accompanied by a partner atom having $q<0$. In order to conserve energy, this
pairing only occurs for $\delta<0$, which allows the potential energy in the molecule to be converted to atomic kinetic
energy for selected modes with $q$ values around $q_{0}=\pm\sqrt{|\delta|}$.

\section{Strategies and future developments}

The stochastic gauge representation is a very general technique suited to a large class of problems. The secret of its success
lies in the possibility of adapting the equations to the particular problem being studied. Already much work has been done
on optimizing the \textbf{gauge} for a variety of problems. Based on this, we conjecture that in the general, a successful
\emph{drift} gauge should a) guarantee inward asymptotic flow, b) generate an attractive manifold on which the gauge itself
vanishes, and c) be used in conjunction with a \emph{diffusion} gauge to minimize sampling error. Using these techniques
we have been able to extend the useful time of purely unitary (lossless) quantum time-evolution by two orders of magnitude
compared to the positive-P method, for anharmonic oscillator time-evolution with up to $10^{10}$ bosons.

Another, complementary strategy is to optimize the \textbf{basis set}. A basis set that incorporates more of the physics
of the problem being studied would lead to more efficient sampling. Thus, we have recently formulated a phase-space technique
that uses a \emph{Gaussian} basis\cite{CornDrum}, which is also suited to generating a phase-space method for \emph{fermions}.

A third strategy is to optimize the stochastic \textbf{sampling}, employing such QCD-like techniques as the \emph{Metropolis}
and \emph{genetic} algorithms. So far, these techniques have only just started to be investigated.

Perhaps the most important overall conclusion is that while ab-initio quantum dynamical simulations are difficult, they are
not ruled out in principle. These techniques can serve as useful benchmarks in establishing the validity of approximate methods.
They are not limited to quantum problems, since they can also be used for kinetic, genetic or chemical master equations.
But the most interesting development will be in new tests of quantum mechanics for systems that are both strongly
entangled and macroscopic - and these are the types of quantum state whose time-evolution is not easily predicted using previous
methods.


\begin{thebibliography}{10}
\bibitem{Feynman}R. P. Feynman, Int.~J.~Theor.~Phys. \textbf{21}, 467 (1982). 
\bibitem{GaugeP}P. Deuar and P. D. Drummond, Phys. Rev. A \textbf{66}, 033812 (2002).
\bibitem{Wineland}D. Kielpinski, C. Monroe and D. J. Wineland, Nature \textbf{417}, 709 ( 2002). In the table we assume error-correction with
$100$ ions per logical qubit, and $10$-fold logical parallelism.
\bibitem{Digital}Estimates of digital integration levels, cost and performance are based on existing engineering studies and industry projections,
and have at least an order of magnitude uncertainty. Clock speeds are rounded down from the 2016 estimates contained in:
\textbf{HTTP://public.itrs.net/}, \emph{International Technology Roadmap for Semiconductors 2002 Update.}
\bibitem{wilson:74}K.~G.~Wilson, Phys.~Rev.~D \textbf{10}, 2445 (1974); D.~M.~Ceperley, Rev.~Mod.~Phys. \textbf{67}, 279 (1995).
\bibitem{Wig-Wigner}E.~P. Wigner, Phys.~Rev. \textbf{40}, 749 (1932). 
\bibitem{Hus-Q}K.~Husimi, Proc.~Phys.~Math.~Soc.~Jpn. \textbf{22}, 264 (1940). 
\bibitem{Gla-P}R.~J.~Glauber, Phys.~Rev. \textbf{131}, 2766 (1963); E.~C.~G.~Sudarshan, Phys.~Rev.~Lett. \textbf{10}, 277 (1963).
\bibitem{+P}S. Chaturvedi, P. D. Drummond and D. F. Walls, J. Phys. A \textbf{10}, L187-192 (1977); P.~D.~Drummond and C.~W.~Gardiner,
J.~Phys.~A \textbf{13}, 2353 (1980). 
\bibitem{carterdrummond:87}S.~J.~Carter, P.~D.~Drummond, M.~D.~Reid, and R.~M.~Shelby, Phys.~Rev.~Lett. \textbf{58}, 1841 (1987); P. D. Drummond
and A. D. Hardman, Europhys. Lett. 21, 279 (1993). 
\bibitem{Solexp}M. Rosenbluh and R. M. Shelby, Phys. Rev. Lett. \textbf{66}, 153 (1991); P.D. Drummond, R. M. Shelby, S. R. Friberg \& Y.
Yamamoto, Nature \textbf{365}, 307 (1993). 
\bibitem{evapcool}P.~D.~Drummond and J.~F.~Corney, Phys.~Rev.~A \textbf{60}, R2661 (1999). 
\bibitem{GGD-Validity}A.~Gilchrist, C.~W.~Gardiner, and P.~D.~Drummond, Phys.~Rev.~A \textbf{55}, 3014 (1997). 
\bibitem{Paris1}I. Carusotto, Y. Castin and J. Dalibard,~Phys. Rev. A \textbf{63}, 023606 (2001); I. Carusotto, Y. Castin, J. Phys. B \textbf{34},
4589 (2001). 
\bibitem{Plimak}L. I. Plimak, M. K. Olsen, M. J. Collett, Phys. Rev. A \textbf{64}, 025801 (2001). 
\bibitem{pdpdcanprl}P. D. Drummond, P. Deuar, K. V. Kheruntsyan, cond-mat/0308219 (2003). 
\bibitem{1Dg2}K. V. Kheruntsyan, D. Gangardt, P. D. Drummond and G. Shlyapnikov, Phys. Rev. Letts. \textbf{91}, 040403 (2003). 
\bibitem{Twinbeams}K. V. Kheruntsyan and P. D. Drummond, Phys. Rev. A \textbf{66}, 031602(R) (2002). 
\bibitem{CornDrum}J. F. Corney and P. D. Drummond, quant-ph/0308064 (2003).\end{thebibliography}
\end{document}